# Coordination Group Formation for Online Coordinated Routing Mechanisms

**Wang Peng**
University of Florida
Gainesville, FL 32611
Tel: 312-610-3823 Email: pengw@ufl.edu

**Lili Du, Corresponding Author**
University of Florida
Gainesville, FL 32611
Tel: 352-294-7805 Email: lilidu@ufl.edu

## ABSTRACT

This study considers that the collective route choices of travelers en route represent a resolution of their competition on network routes. Well understanding this competition and coordinating their route choices help mitigate urban traffic congestion. Even though existing studies have developed such mechanisms (e.g., the CRM [1]), we still lack the quantitative method to evaluate the coordination penitential and identify proper coordination groups (CG) to implement the CRM. Thus, they hit prohibitive computing difficulty when implemented with many opt-in travelers. Motived by this view, this study develops mathematical approaches to quantify the coordination potential between two and among multiple travelers. Next, we develop the adaptive centroid-based clustering algorithm (ACCA), which splits travelers en route in a local network into CGs, each with proper size and strong coordination potential. Moreover, the ACCA is statistically secured to stop at a local optimal clustering solution, which balances the inner-cluster and inter-cluster coordination potential. It can be implemented by parallel computation to accelerate its computing efficiency. Furthermore, we propose a clustering based coordinated routing mechanism (CB-CRM), which implements a CRM on each individual CG. The numerical experiments built upon both Sioux Falls and Hardee city networks show that the ACCA works efficiently to form proper coordination groups so that as compared to the CRM, the CB-CRM significantly improves computation efficiency with minor system performance loss in a large network. This merit becomes more apparent under high penetration and congested traffic condition. Last, the experiments validate the good features of the ACCA as well as the value of implementing parallel computation.

## 1. Introduction

Travelers driving en route usually act strategically in an effort to choose the best routes so that they can achieve the safest and most efficient journey possible. When a group of travelers simultaneously make these route choices, which make them potentially pass the same roads in the same time interval, they are effectively in a competition for network route resources. We use an extreme example shown in Fig. 1 to demonstrate this view, assuming the capacity of each colored link is one and each traveler has two candidate routes, such as routes $i_1$ and $i_2$ for traveler $i$. The network shows that the routes of travelers i and k overlap, so do the routes of travelers k and j. Given the route capacity is one, the route choices of traveler i and traveler $k$ will affect each other, so as traveler $k$ and traveler $j$. At the meantime, it is noticed that the route choices of travelers $i$ and $j$ can also affect each other indirectly through traveler $k$. This example clearly demonstrates that the route choice of individual travelers presents their competition on using network route resources.

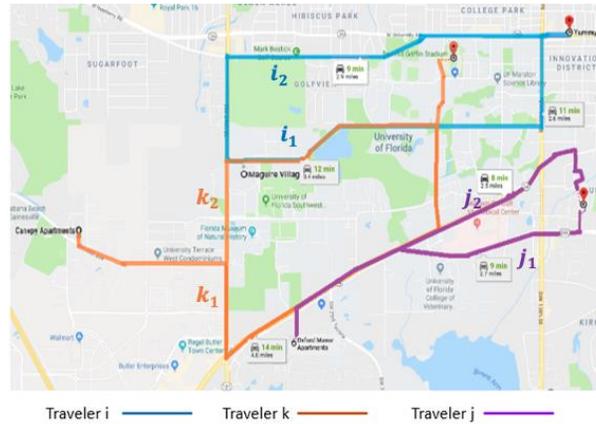

Fig. 1. Example of three travelers competing for traffic resources.

Well understanding and coordinating such competition on using route resources in a traffic network will help mitigate urban traffic congestion. Many previous studies and emerging technologies have already been and will be developed to guide route choices of traffic demands toward an optimal solution, which either mainly values a traveler's optimality (such as independent in-vehicle routing mechanism (IRM)) or the traffic system's performance (such as congestion pricing strategies). However, it has been recognized that IRM, by which each traveler independently conducts the best response to the real-time traffic condition, will result in traffic congestion oscillation among multiple candidate routes over traffic network, and eventually jeopardize both system and user traffic efficiency [14] while congestion pricing strategies are not well accepted by travelers since they increase the travel costs. To be noted, the strategy of IRM is implemented for each individual traveler, while congestion pricing strategy is systematically implemented for all travelers. To address the above dilemma, the recent studies developed by [1] propose coordinated online in-vehicle routing mechanism (CRM), which seeks to balance individual user and system optimality without violating selfish nature of individual traveler. Briefly, the CRM works on an online route coordination group, in which each traveler iteratively proposes and updates its routing choice priorities, responding to real-time traffic information (incorporating their most recent route choices) through connected vehicle environment; the negotiation process ends with an equilibrium route choice decision among all smart travelers, which lead to a better overall system performance compared to IRM.

However, to implement the CRM efficiently, we face the immediate difficulty of identifying the coordination group. More exactly, the CRM will lose its value if it works on a small coordination group with weak coordination potential among travelers, but will meet prohibitive computation load for a coordination group involving many travelers with a very complicated competition on network resources. Therefore, it is critically important to understand the online competition among travelers and identify proper coordination groups in terms of both coordination potential and group size. Note that the competition potential among travelers on network route resources demonstrates the potential opportunity to coordinate their route choices (i.e., coordination potential) for traffic congestion management.

On the other hand, it is very difficult to identify this competition among multiple travelers driving en route since it involves many complicated temporal-spatial factors. For example, the candidate routes of the two travelers (such as travelers $i$ and $k$ in Fig. 1) which have a high level of spatial (link) overlap may have high potential to compete for these common road resources. However, if they pass the overlapped links at different time intervals, then they do not compete on the usage of these routes. Moreover, two travelers (such as traveler $i$ and $j$ in Fig. 1), whose candidate routes don't spatially overlap, may each intend to use (i.e. competes on) the roads overlapping with the same third traveler (such as traveler $k$ Fig. 1), and then they still compete indirectly. With this observation, this study names the competition potential between two travelers like $i$ and $k$ as the direct competition potential, while considers the competition potential between two travelers like $i$ and $j$ as the indirect competition potential. Moreover, it is recognized that the competition potential holds non-additive features. Namely, the indirect competition potential between travelers $i$ and $j$ is not equal, but weaker to the sum of the direct competition potential of travelers $i$ and $k$, and travelers $j$ and $k$. The indirect competition between two travelers may go through different intermediate travelers as we extend the concept to multiple (>3) travelers. These properties indicate that the competition potential is not in Euclidean space. We still lack of methodology to quantify and evaluate the competition potential (i.e. coordination potential) among all travelers en route in a big traffic network. As a result, we can neither identify a coordination group of travelers with strong coordination potential, nor efficiently coordinate their online route decisions to mitigate traffic congestions.

This study seeks to make up the above technology gap. Specifically, we consider a group of travelers driving en route and develop quantitative approaches to better understand the competition potential among travelers according to their current OD and candidate route sets. Built upon that, we will design a machine learning approach to cluster the coordination groups with strong competition/coordination potential, which can be used to support the implementation of coordinated in-vehicle online routing mechanisms. To conduct this research, we assume the availability of online OD demand data, which have been well collected from multiple sources such as GPS and pickup/drop data from shared mobility services such as Uber and Lyft. Below presents the main research contributes along with the research efforts.

First of all, by taking account of the temporal-spatial overlap of the candidate routes of two travelers as well as the travel time uncertainty of a traffic network, this study contributes multiple mathematical formulations to capture the direct competition potential between two travelers. Next, to present the indirect competition in a group of travelers of interests, this study establishes a competition network where a node represents a traveler, a link represents a direct competition between two travelers and the weight of a link is assigned as the reciprocal of the direct competition potential. Considering the features of the competition potential, this study quantifies the indirect competition potential between two travelers by the reciprocal of the distance of the shortest path on the competition network. Built upon the understanding of the direction and indirect competition through the competition network, this study introduces the concentration to quantify the competition potential within a group of travelers (i.e., inner-cluster competition).

Furthermore, it is recognized that some travelers present strong competition potential but others not. Coordinating the route choices of all travelers driving en route in a big network leads to prohibitive computation load. This study, therefore, develops an adaptive centroid-based clustering algorithm (ACCA) to break a large group of travelers into multiple smaller coordination groups. Along with the development of the ACCA, it is recognized that for a given traveler group, generating more clusters each with a smaller size often leads to a higher inner-cluster competition (the competition potential among travelers in a group) but also more complicated inter-cluster competition (the competition potential among travelers in different groups). While inner-cluster competition promotes the function of the coordinated online traffic management, the inter-cluster competition represents the interference to jeopardize the performance. In fact, this dilemma represents a general tricky for the graph clustering algorithm. There isn't a universal solution ([11]). According to the characteristics of the problem, this study contributes the new formulations to quantify the benefit and cost from including one more cluster in the ACCA and then considers the marginal net benefit as a reference to locate a local optimal clustering solution. The local optimality of this good clustering solution is ensured by the statistical analysis. The algorithm analysis shows that the ACCA algorithm itself causes computation obstacles for real-time application. The computation process of competition network construction and competition relationship identification requires to go through every pair of travelers, and during the clustering process, the assignment of each node to its centroid and the update of the centroid are carried in each iteration. And these procedures are key factors that affect the computation efficiency of ACCA. Thus, computation load analysis is carried in section 4.4.2, 4.4.3 and mainly in section 4.6. And because of the independence of computation process in these procedures, we implement the ACCA by parallel compuation to improve the computation efficiency. Last, this study proposes a clustering based coordinated routing mechanism (CB-CRM), which implements the CRM on individual clusters and coordinates their route choices in each cluster.

The applicability of the approaches contributed by this study is evaluated by the numerical experiments on Sioux Falls city network and Hardee city network. The experiments use the IRM and the CRM implemented on an entire traveler set as the benchmarks to demonstrate the merits of the CB-CRM in both improving system performance and reducing computation load.

Specifically, the experiments first validate the applicability of the ACCA for this online application. It is noticed that clustering travelers in a traffic network through the ACCA still introduces extra computation load beyond the computation load of the CRM, even though the parallel computation helps improve the computation efficiency of the ACCA significantly. Thus, the CB-CRM fits better to a large network with a large scale travelers, which are separated into multiple coordination groups. Next, the stop criterion of the ACCA adapts to both separable and non-separable data. It is able to drive the ACCA to stop at a local optimal number of clusters, from which the system performance has a significant drop. Moreover, after comparing different formulations for quantifying the direct competition potential, we noticed that the formulation considering temporal-spatial link usage overlap provides the best performance. Furthermore, the experiments demonstrate that the CB-CRM outperforms IRM in system performance (i.e., resulting in a smaller system travel cost). The benefit becomes more significant under higher penetration and more congested traffic state. Thus, we conclude that the CB-CRM is able to greatly alleviate traffic congestion. As it's compared to the CRM, the CB-CRM demonstrates minor system performance loss to gain computation efficiency, especially in a large network like Hardee network. In addition, our experiments demonstrate that the CB-CRM using the ACCA results in less traffic congestion than the CB-CRM built upon RCA. Therefore, the ACCA works effectively to cluster travelers with strong competition potential which leads to a high potential to be coordinated.

Overall, this study claims that it is very valuable to understand and quantify the competition potential among travelers driving en route so that we can setup efficient coordinated online traffic management schemes to mitigate network-level congestion without involving prohibitive computation load. To the best of our best knowledge, this study is among the first attempt to investigate the knowledge and methodology to quantify the route competition/coordination potential among travelers en route, and the associated schemes for mitigating network level traffic congestion in the transportation field. This study can be further extended to other applications such as parking and riding sharing, in which users also present the potential competition or cooperation, and coordinated management is applicable for better system performance. Thus, this study will contribute both the literature as well as promotes the development of online traffic control and management.

The research efforts are presented by the following structure. We first briefly review the most related research in literature and then present the problem statement. Built upon that, we provide the methodologies to quantify the direct and indirect competition potential between any two travelers or multiple travelers in a network. Furthermore, the ACCA is designed to separate a large scale of travelers into small coordination groups. Last, numerical experiments are set up to validate our study. The paper ends with a summary of the research contributions and future work.

## 2. Literature review

The research problem initiated in this paper is quite innovative and there is not much existing work for the references. From the perspective of modeling traveler cluster, the most related work is the traffic analysis zone (TAZ) used in transportation planning models [13]. Nevertheless, TAZs mainly reflects the land-use and the population distribution rather than the detailed route resources competition among travelers driving en routes. Similarly, the mode split used in transportation planning models [13] address the competition between travelers on different traffic modes, which is also different from this study.

From the perspective of the solution approaches, the ACCA designed in this study is related to the design of graph clustering algorithm in computer sciences. Considering there are many various algorithms, the brief review below will mainly focus on the mainstream algorithms in literature and demonstrate the inapplicability of these algorithms to our study, considering the two critical characteristics of this particular application. (i) The data is not in Euclidean space. (ii) The target number of the clusters is unknown, but with a great concern for this application. Along with this discussion, we justify the unique contributions of the ACCA as well.

First of all, we rule out the algorithms of K-means [3], mean shift, and K-medoids [10] for this study since they work for the data in Euclidean space and need to predetermine the number of clusters in the clustering solution. Next, both Affinity propagation [8] and Spectral clustering algorithms [4] work for non-Euclidean space data, but none of them fit this study well since Affinity propagation algorithm cannot ensure to end with the good number of clusters, which balances the inner- and inter- cluster competition, and the procedure of the Spectral clustering algorithm involves a K-means algorithm which still requires a pre-determined target number of clusters. Last, DBSCAN [7] uses density-connected condition to identify the border of high-density areas, and only recognizes the nodes with high-density but ignores other nodes outside. Clearly, DBSCAN does not fit this study since there are no nodes (travelers) should be ignored in this study.

We next review some clustering methods which particularly address graphical data. Minimum spanning tree [5] (MST) based clustering algorithm finds the MST of the original graph first, and then break an edge at each iteration so that the algorithm generates one new cluster represented by the new generated subtree. Again, such algorithm does not fit this study well since it cannot ensure a solution well balancing the inner- and inter- cluster competition. Hierarchical clustering [6] is separated into two categories including divisive and agglomerative clustering. Hierarchical divisive clustering is similar to the MST which generates new smaller clusters by separating an original cluster. And Hierarchical agglomerative clustering is the opposite way that recursively merges a pair of clusters as one cluster. Both of them need to predetermine the number of clusters in the solution. Therefore, they are not suitable for our problem.

The above review indicates the need to develop a customized clustering algorithm to satisfy the particular applications in this study. The ACCA algorithm works on the competition network and seeks to break the set of nodes (travelers driving en route) into different clusters so that travelers in each cluster present strong competition potential. Moreover, the ACCA ensure the algorithm to end at a local optimal clustering solution so that we can balance the inner- and inter- cluster competition. Those good qualities demonstrate the unique contribution of the ACCA.

## 3. Problem statement

This study considers a group of travelers driving en route on a city network and investigates the competition potential among the travelers on the usage of route resources in this traffic network. Specifically, this study defines the competition potential (CP) among travelers as the potential that the travelers will use the same traffic roads in a future time interval. The proposed research efforts seek to well understand and quantify this competition so that we can coordinate their online route choices to mitigate traffic congestion. To address this research need, this study develops mathematical approaches to quantify the direct and indirect CP between two travelers and further identifies coordination groups in which the travelers present strong CP. To do that, we formulate the problem as follows. The city network is considered as directed network with node set $N$ and link set $A$. The indices $n \in N$ and $a \in A$ are used to represent individual nodes and links, respectively. Each link in the network is associated with a travel time $t_a$, which follows a known distribution with mean and variance denoted by $\bar{t}_a$ and $\sigma_a^2$, respectively. The group of the travelers, differentiated by their ODs, is denoted by $I$ (also denote the maximum number of travelers). Each traveler $i \in I$ has $i_H$ candidate routes linking its OD. We further introduce $i_h$ to represent the $h^{th}$ candidate route of traveler $i$, $h = 1,2, \ldots, H$. $a_{i_h}$ represents the link set of the $h^{th}$ route of traveler $i$, which is composed of $|a_{i_h}|$ number of links. The average route travel time can be calculated by $T_{i_h} = \sum_{a \in a_{i_h}} \bar{t}_a$. As mentioned in the introduction section, both direct and indirect competition will happen between two travelers on the traffic network. Accordingly, for any two travelers $i, j$, we introduce $c_{ij}$ and $c_{i \to j}$ to denote their direct and indirect competition potential respectively.

The critical research challenge here is how to quantify the competition potential between two travelers as well as among multiple travelers for using the network route resources considering temporal-spatial traffic variation. In addition, the computation efficiency is of particular concern since the understanding of this CP is going to be used to support online traffic management, such as coordinating real-time route choices of a group of vehicles with high CP. This study addresses these research difficulties in the next section.

## 4. Methodology

This section first develops mathematical formulations to measure the direct CP between two travelers according to the temporal-spatial potential overlap among their candidate routes. Next, we build a competition network, from which we develop graphical approaches to identify the indirect CP between two travelers as well as a group of travelers. Last, this study designs an adaptive centroid-based clustering algorithm to cluster a large scale of the travelers into multiple coordination groups, in each of which the travelers present a strong competition potential on the usage of the network route resources. The efficiency and applicability of the proposed approaches are evaluated by a clustering based coordinated online in-vehicle routing mechanism implemented on Sioux Falls city and Hardee network.

### 4.1 Direct competition potential between two travelers

This section develops the formulations to quantify the direct CP between two travelers. We consider that the overlap between their respective candidate routes represents the potential that these two traveler will compete on the usage when they make route choices en route. Clearly, a higher level of the overlap represents a stronger competition potential. Accordingly, Eq. (1)) and Eq. (2) respectively measure the level of the potential overlap among the candidate routes of the two travelers from the spatial dimension without counting the competition possibility in the temporal dimension.

$$c_{ij} = \sum_{j_h} \sum_{i_h} \left( \frac{\sum_{a \in a_{i_h} \cap a_{j_h}} o_a}{|a_{i_h}| + |a_{j_h}|} \right) \tag{1}$$

$$\text{or } c_{ij} = \sum_{j_h} \sum_{i_h} \left( \frac{\sum_{a \in a_{i_h} \cap a_{j_h}} \bar{t}_a o_a}{\sum_{a \in a_{i_h}} \bar{t}_a + \sum_{a \in a_{j_h}} \bar{t}_a} \right), \tag{2}$$

where for two given travelers $i$ and $j$, we set $o_a = 1$, if $a \in a_{i_h}$ and $a \in a_{j_h}$; otherwise $o_a = 0$.

It is recognized that two travelers may not compete on using a spatial overlapped road if they pass the road at different time intervals. This study next develops a measurement considering the competition possibility in temporal-spatial space. Specifically, the temporal-spatial overlap among their candidate route sets are only counted if these two travelers likely traverse the same link in the same time interval. Namely, there should have a temporal overlap between the arrival time of one travelers and the departure time of the other traveler at a spatial overlapped link. Following this idea, this study introduces Eq. (3) below to quantify the time-dependent direct CP between two travelers.

$$c_{ij} = \sum_{j_h} \sum_{i_h} \left( \frac{\sum_{a \in a_{i_h} \cap a_{j_h}} \theta_a \bar{t}_a o_a}{\sum_{a \in a_{i_h}} \bar{t}_a + \sum_{a \in a_{j_h}} \bar{t}_a} \right) \tag{3}$$

where the indicator $\theta_a$ is introduced to take account of the temporal possibility that travelers $i$ and $j$ will compete for the usage of a road in the same time interval. Note that the value of $\theta_a$ can be identified by Eq. (4) considering average link/route travel time or Eq.(5) which is further refined by considering the link travel time uncertainty.

$$\theta_{a1} = \begin{cases} 1, & if \left(\overline{T}_{a_j} \leq \overline{T}_{a_i} \leq \overline{T}_{(a+1)_j}\right) or \left(\overline{T}_{a_i} \leq \overline{T}_{a_j} \leq \overline{T}_{(a+1)_i}\right) ; \forall a \in a_{i_h} \cap a_{j_h}, i, j \in I \\ 0, & otherwise \end{cases} \quad (4)$$

where $T_{a_i}$ represents the arrival time of traveler $i$ at the overlapped road, i.e., $\overline{T}_{a_i} = \sum_{l=1}^{a} \bar{t}_{l_i}$.

$$\text{or } \theta_{a2} = P\left(\overline{T}_{a_j} \leq \overline{T}_{a_i} \leq \overline{T}_{(a+1)_j}\right) + P\left(\overline{T}_{a_i} \leq \overline{T}_{a_j} \leq \overline{T}_{(a+1)_i}\right), \forall a \in a_{i_h} \cap a_{j_h}, i, j \in I \quad (5)$$

where $T_{a_i} = \sum_{l=1}^{a} t_{l_i}$ is a random variable representing the random arrival time of traveler $i$ at the overlapped road. To calculate $\theta_{a2}$, this study considers that the link travel time $t_a, a \in a_{i_h}$ follows independent Normal distribution (i.e., $t_a \sim N(\bar{t}_a, \sigma_a^2)$) so that we have $T_{a_i} \sim N(\sum_{l=1}^{a} \bar{t}_{l_i}, \sum_{l=1}^{a} \sigma_{l_i}^2)$. Two reasons support us to use Normal distribution. First, there is existing literature [2] which shows the travel time in shorter trips tended to be distributed normally while in longer trips it follows a log-normal distribution. Second, it has been shown that most of the other distribution can be approximated by normal distribution [17]. As the link flow interdependence is considered, more complicated statistical tools such as correlation matrix will be used to capture $\theta_{a2}$. Overall, Eq. (1)) and (2) only considers the potential overlap of candidate routes from the spatial dimension. Eq. (3) together with Eq. (4) considers the potential overlap of candidate routes from the temporal-spatial space, and Eq. (3) together with Eq. (5) further takes account of the uncertainty of travel time. Our numerical experiments show the temporal-spatial formulations involve better traffic reality than spatial formulations. But this advantage comes with additional computational complexity. We may need to balance the model accuracy and computation load for online applications.

### 4.2 Indirect competition potential among multiple travelers in competition network

Indirect competition happens between any two travelers when their candidate route sets do not have spatial overlap, but both compete for the usage of the network resources with the same third traveler (e.g. traveler $i$ and $j$ in Fig. 1). This type of indirect competition may occur in a very complicated manner among multiple travelers on a traffic network. To study this indirect competition potential among multiple travelers (>3), this study builds up a competition network $\Gamma(I, L)$. An example is shown in Fig. 2, in which each node indicates a traveler and each link indicates the existing of a direct competition potential. We will introduce the weight of the link later in this section. Built upon that, this study next develops a graphical approach to quantify the indirect competition potential.

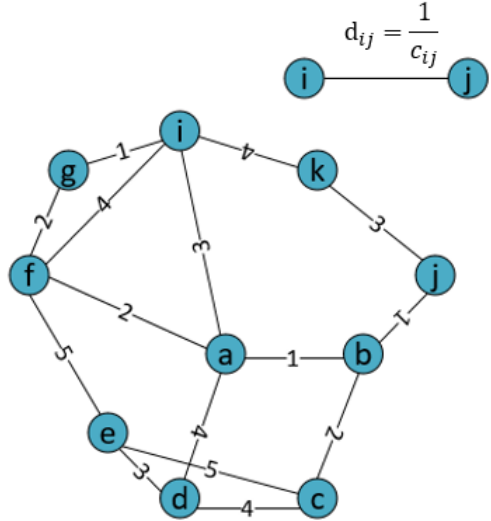

Fig. 2 Competition Network $\Gamma$(I,L)

To specify the weights in the competition network, we first discuss the fundamental properties of indirect CP, using the example shown in Fig. 2. First of all, it is noticed that the indirect competition between two travelers may go through multiple possible and very complicated pathway involving multiple intermediate travelers. For example, traveler $g$ and $b$ may indirectly compete through travelers $f$ and $a$ (see the pathway: $g - f - a - b$) or through travelers $i, k, j$ (see the pathway: $g - i - k - j - b$). Next, the competition potential is non-additive. Namely, if two travelers $i$ and $j$ only compete through the third traveler $k$, then $c_{i \to j} < c_{ik} + c_{jk}$. However, a larger value of $c_{ik} + c_{jk}$ indicates a stronger indirect competition $c_{i \to j}$. According to the above properties, we introduce the formulation in Eq. (6) to measure the indirect CP of two travelers involved in a competition network.

$$c_{i \to j} = max \left\{ 1 \middle/ \sum_{uv \in L(i \to j)} \frac{1}{c_{uv}}, \forall L(i \to j) \in \Gamma \right\}, \forall i, j \in I \qquad (6)$$

where $L(i \to j)$ represents a pathway from traveler $i$ to traveler $j$ on the competition network $\Gamma$. Clearly, Eq.(6) satisfies the non-additive property and implies $c_{i \to j} = c_{ij}$ if travelers $i$ and $j$ directly compete for the network route resource. Therefore, Eq.(6) is able to measure both direct and indirect CP between two travelers. Moreover, Eq.(6) uses the pathway with the largest value of $1/\sum_{uv \in L(i \to j)} \frac{1}{c_{uv}}$ to measure the indirect competition potential, which corresponds to a pathway involving the strongest direct competition potential between travelers along the pathway.

Even though $c_{i \to j}$ defined in Eq.(6) has several good properties, it is hard to be calculated on the competition network $\Gamma$. However, its reciprocal (denoted as $d_{ij}$) in Eq. (7) demonstrates a better network calculation feature. For example, $d_{ij}$ can be easily found by searching the shortest path, if we assign the weight of the link in the competition network $\Gamma(I, L)$ as the reciprocal of the direct competition potential of two travelers. To this end, we solve the issue of assigning the weights of the competition network as well as quantifying the indirect competition between any two travelers.

$$d_{ij} = \frac{1}{c_{i \to j}} = min \left\{ \sum_{uv \in L(i \to j)} \frac{1}{c_{uv}}, \forall L(i \to j) \in \Gamma \right\}, \forall i, j \in I \qquad (7)$$

Note that the value of the $d_{ij}$ between any two nodes $i$ and $j$ on the competition network $\Gamma(I, L)$ represents the distance of the shortest path between these two nodes. This study will use both the terms of node and traveler in the rest of the paper to make the articulation fit the context.

### 4.3 Competition potential between among multiple travelers

Built upon the competition network, this study further develops the mathematical measurement to quantify the competition potential among a group of travelers driving en route. To do that, we first introduce the concept of the centroid of a cluster.

*Centroid of a cluster*: A node in $\Gamma(I, L)$ is defined as the centroid of a cluster if the sum of the distance (weight) from this node to all other nodes within the cluster is the minimal. The centroid set for a clustering solution $G(K)$ is denoted as $C(K) = \{ c_k \}_{k=1}^{K}$, where $c_k$ represents the centroid of the cluster $k$. Since each centroid corresponds to an individual cluster, $c_k$ is also used to represent a cluster within a proper context.

Along with this definition, a cluster, whose centroid is the nearest one from the centroids set to a node in this cluster, has the strongest competition potential to this node among all other existing clusters. Accordingly, we introduce the concept of concentration to measure the competition potential among travelers within a cluster (i.e., inner-cluster competition potential).

*Concentration*: For a cluster with a centroid $c_k$, its concentration is calculated by Eq. (8).

$$s_k = \sum_{i=1}^{I} \gamma_{ik} d_{ic_k}, k \in K \qquad (8)$$

where $\gamma_{ik} = 1$, if node $i$ belongs to the cluster k; otherwise $\gamma_{ik} = 0$. $d_{ic_k}$ represents the distance from node $i$ to the centroid $c_k$. Clearly, a cluster with a smaller concentration presents stronger competition potential. Note that the concept of the concentration in this study is different to the concentration used in K-centered algorithm (i.e., K-medoids [10] and K-means [3] in literature where it is defined in a Euclidean space.

### 4.4 Identify coordination group through ACCA clustering algorithm

It is recognized that some travelers en route present strong competition potential but others not. The coordinated traffic management and control schemes (such as the CRM developed by [1]) often have interests in guiding the route decisions of travelers with strong competition potential. At the meantime, coordinating the route choices of a large scale of travelers en route in a big city will cause prohibitive computation load. This section thus thinks of clustering the travelers into different coordination groups according to their competition potential.

Regarding this particular application, the algorithm needs to address another critical challenge. It is the trade-off between the number of clusters and the size of each cluster. Specifically, we expect a small cluster with a strong inner-cluster competition potential so that a small computation load is obtained. However, keeping a small size for each cluster will lead to a large number of clusters for a given traveler group of interests. Given it is hard to make inter-cluster competition (i.e., the CP between two clusters) zero, a large number of clusters will lead to complicated inter-cluster CP, which is the interference having a negative effect on the performance of the coordinated traffic management and control. The clustering algorithm design in this study will address the above tradeoff by exploring an optimal set of clusters in the solution.

### 4.4.1 Adaptive Centroid-based Clustering Algorithm

This section designs the Adaptive Centroid-based Clustering Algorithm (ACCA) to cluster travelers driving en route into multiple small coordination groups while balancing the inner- and inter-CP. The procedure of the algorithm is given below. Followed that, we explain the main ideas and the technical details.

Step 0: Initialization: Competition Network $\Gamma(I, L)$; calculate the indirect CP

Step 1: Set K=1, and select centroid $c_1$; set the cluster set $G(1)$

Step 2: Exploring optimal clustering solution

    1) K=K+1

    2) Generate the initial centroids set $C(K)$, which maximizes the concentration reduction $b_n$ in Eq. (9);

    3) Explore clustering solution $G(K)$

        a) Assign each node $i$ to its nearest centroids $c_k$, $\forall\, c_k \in C(K)$;

        b) Update the centroid set $C(K)$;

        c) Repeat (a) and (b) until reaching convergence;

        d) Update $G(K)$;

    4) Evaluate $G(K)$ by criterion $r(K)$ defined in Eq. (13);

    5) If $r(K^*)$ is the maximum of { $r(K)$ } in consecutive five times, then go to Step 3, otherwise go to step 2.1

Step 3: Stop and return $K^*$

Mainly, the ACCA takes all travelers as one cluster at the beginning and then breaks it into multiple clusters (adding one more in each iteration) until an optimal set of clusters is reached. The procedure to explore a new clustering solution includes the key operations: (1) generating new initial centroid set; (2) assigning each node to its 'nearest' centroid; (3) refine the centroid for each cluster. And the procedures of (2) and (3) repeat multiple iterations until the concentration of the cluster solution converges. The splitting process stops if the clustering solution satisfies the optimal criterion. The ACCA involves two technical challenges (i) how to determine an optimal clustering solution regarding the aforementioned tradeoff; (ii) how to find a good initial centroids set. We discuss the computation complexity and also address the two design difficulties in following sections.

### 4.4.2 Exploring clustering solution

With the initial centroids set, the first operation is to assign each node $i$ to its nearest centroids $c_k$. Mathematically, it's to find the minimum distance from each node to the centroids set. It is done by the algorithm of finding minimum value in an array. Finding a node $i$'s nearest centroid in a centroids set of size $K$ requires $(K-1)$ comparison between the current nearest centroid and every other centroid in this array, then its time complexity is O(K). So that the time complexity of this whole procedure is $O(KI)$ because this operation needs to repeat for every node, and $I$ is the node set of competition network but here it also denotes the total number of nodes (travelers) on the competition network.

And the second operation is to refine the new centroid for each new cluster obtained from the first operation. This requires two steps: 1) summation of the distance from each node to all other nodes within its cluster; 2) find the node with minimum sum which will be the new centroid. The first step for a cluster $c_k$ of size $|c_k|$ requires $(|c_k|-1) \cdot |c_k|$ computations, then its time complexity is O($|c_k|^2$). The second step is similar to the first operation, and its time complexity for a cluster $c_k$ is O($|c_k|$).Thus, the time complexity of the second operation is O($\sum_{k \in K} |c_k|^2 + |c_k|$). And since $\sum_{k \in K} |c_k| = I$ and the property$(a+b)^2 \geq (a^2+b^2)$, thus the time complexity of procedure (3) is bounded by O($I^2 + I$).

### 4.4.3 Searching good initial centroid sets

Similar to K-means and K-medoids algorithms, and the solution of the ACCA is sensitive to the initial centroids too. Existing literature provides naive approaches, including trying out all points or try $h$ (such as $h = 100$) number of sampling points. They usually either cause high computation load or no goodness is guaranteed. This study thus develops a heuristic algorithm to search a good initial centroid set in a more strategic way, when a new cluster is added to the solution. The main idea is described as follows.

With current centroid set $C(K-1)$, we generate a new centroid set $C(K)$ by looking for a new centroid $c_K^0$ which minimizes the concentration of the new clustering solution with $K$ clusters, while keeping $C(K-1)$ unchanged. Namely, the new centroid set is formed by $C(K) = (C(K-1), c_K^0)$. To do this, we define the concentration reduction by adding node $c_K \in I$ as the new centroid through $\psi_{c_K} = \sum_{i=1}^{I} \max(d_{i\hat{c}} - d_{ic_K}, 0)$, in which $d_{i\hat{c}}$ is the distance between node $i$ and its nearest centroid $\hat{c}$ (i.e., node $i$ belong to the cluster $\hat{c}$ in the solution $G(K-1)$) ; and $d_{ic_K}$ denotes the distance between node $i$ and the new centroid node $c_K$. Then, the best new initial centroid for the new added cluster is identified by Eq. (9) below.

$$c_K^0 = \arg\max_{c_K \in I}\{\psi_{c_K}\} = \arg\max_{c_K \in I}\left\{\sum_{i=1}^{I} \max(d_{i\hat{c}} - d_{ic_K}, 0)\right\} \tag{9}$$

This heuristic algorithm helps us strategically find a new good initial centroid set. However, Eq. (9) indicates that we need to go through every node ($I$ in total) in the computation network to measure a $\psi_{c_K}$ and then do the same calculation for every candidate centroid node $c_K$ (i.e., every node in the competition network, thus has $I$ number of nodes in total). Accordingly, Eq. (9) involves a computation complexity of $O(I^2)$ and it has to be re-calculated every time when a new cluster is added to the current solution. This high computation load is not adaptive to the online application. To address this computation issue, this study introduces Eq. (10) to evaluate the potential of the new centroid to reduce the concentration $S_K$. It has been shown that the smaller the value of $w_i$ is, the more likely the node, if used as a centroid of a cluster, will lead to a smaller $S_K$ [12]. Taking advantage of this feature, this study sorts all nodes in competition network $\Gamma(I, L)$ according to their measurements in Eq. (10), and then picks the best node as the new initial centroid if it results in the smallest concentration. Although Eq. (10) and the sorting algorithm respectively involves a computation load of O ($I^2$) and $O(Ilog(I))$ in the modified algorithm, they are only conducted once at the beginning of the ACCA. Thus, the modified algorithm efficiently reduces the computation load.

$$w_i = \sum_{j=1}^{I} \frac{d_{ij}}{\sum_{l=1}^{i} d_{jl}}, i \in I \tag{10}$$

Moreover, the algorithm to search good initial centroid node ensures $S_K \leq S_{K-1} - \max_{i \in I}\{\psi_i\}$, where $S_K = \sum_{k=1}^{K} s_k$ represents the concentration of the clustering solution. Namely, the procedure monotonically reduces the concentration of the clustering solution ($S_K$) as more clusters are involved.

### 4.4.4 Exploring local optimal clustering solution

As the algorithm goes on, the size of each cluster decreases which leads to a smaller value of the cluster concentration (i.e., a stronger inner-cluster CP), while the number of the clusters increases which leads to a higher inter-cluster CP. Thus, the concentration cannot be used as the only reference to evaluate the quality of the clustering solution in this study. This dilemma calls for a new reference to identify a good clustering solution. To address this issue, this study develops the mathematical formulations to quantify the benefit and cost of adding one more cluster to the existing clustering solution. The marginal net benefit is used as the reference to evaluate the quality of a clustering solution.

Specifically, this study defines the benefit of including more clusters as the amount of the reduction of the concentration of the clustering solution, using the concentration of the solution $S_1$ including all travelers in one cluster as the benchmark. Considering $S_1$ has the maximum concentration, this study quantifies the normalized benefit by including K number of clusters in the solution by Eq. (11).

$$r_b(\text{K}) = \frac{S_1 - S_K}{S_1} \tag{11}$$

Next, we consider the sum of the direct CP among travelers in different clusters as a measurement to quantify the inter-cluster CP. Accordingly, this study defines the cost of including more clusters as the increment of the inter-cluster CP. This inter-cluster CP starts with zero and reaches the maximum value when the number of clusters is equal to the number of travelers. Using the sum of the direct CP among all travelers as the benchmark, this study defines Eq. (12) to quantify the normalized cost if K clusters are included in the clustering solution.

$$r_c(\text{K}) = \frac{\sum_{u \notin g(k), v \in g(k), k \in K} c_{uv}}{\sum_{u,v \in \Gamma} c_{uv}} \tag{12}$$

The ACCA expects to stop at a point which leads to the maximum net benefit. Then, the optimal criterion is defined by Eq. (13) below. To implement this criterion, we make the ACCA stop at a local optimal $\text{K}^*$ if $r(\text{K}^*) \geq r(\text{K}) \; \forall \text{K} \in (0, \text{K}^* + 5)$.

$$r(\text{K}^*) = \max\{r_b(K) - r_c(\text{K})\} \tag{13}$$

Furthermore, we demonstrate the existing of such local optimal solution corresponding to this criterion through the statistical analysis. It is noticed that both $r_b(\text{K})$ and $r_c(\text{K})$ starts with value 0 and ends with value 1 as the value of $K$ changes from 1 to $I$ (see Fig. 3 (a)). Accordingly, this study noticed that $r(\text{K}) = r_b(\text{K}) - r_c(\text{K})$ presents a pseudo concavity feature as shown in Fig. 3 (b) (also Fig. 5). To demonstrate the generality of the observation, this study runs 15 different sets of experiments built on Sioux network. Each experiment randomly picks 200-250 travelers with different OD pairs. Using the data $\Delta r(\text{K}) = r(\text{K}) - r(K-1)$ vs. K (see Fig. 4), this study found the inverse regression model: $\Delta r(\text{K}) = \frac{0.218}{K} - 0.006$. The regression model indicates that if $r(K)$ is approximated by a continuous and differentiable function, its second order derivative $r''(\text{K}) < 0$. Thus, these statistical results demonstrate the concavity feature of the $r(\text{K})$. Moreover, the regression model shows that there exists a $\text{K}^* > 0$, where $\Delta r(\text{K}) = 0$, if $K = \text{K}^*$; $\Delta r(\text{K}) > 0$, if $K < \text{K}^*$; and $\Delta r(\text{K}) < 0$, if $K > \text{K}^*$. This further indicates that $\text{K}^*$ represents a local optimal solution that balances the trade-off between inner-cluster CP and inter-cluster CP. This same pattern is also found in the experiments built on Hardee Williston, FL network which is consist of 44 nodes and 134 links.

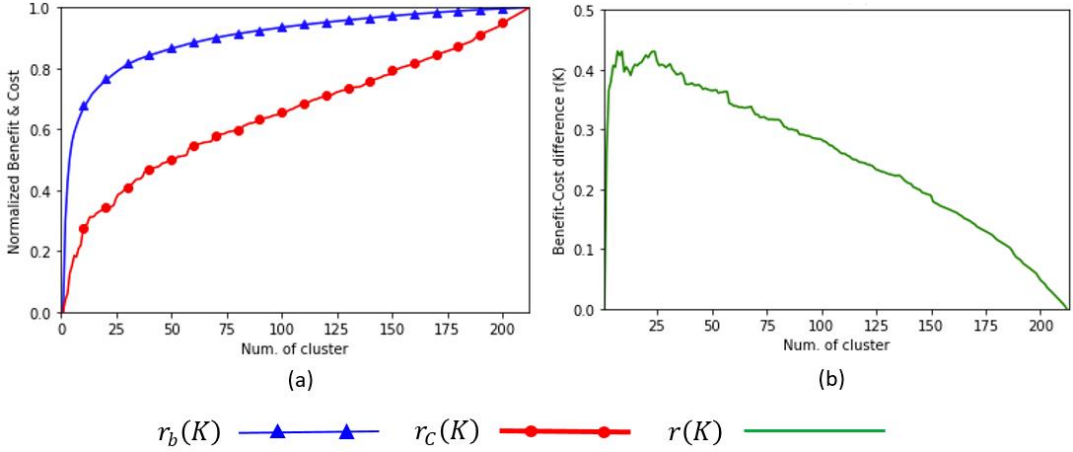

Fig. 3 The $r_b(K)$ & $r_c(K)$ (a) and $r(K)$ (b) wrt. the number of clusters (200 travelers, Sioux)

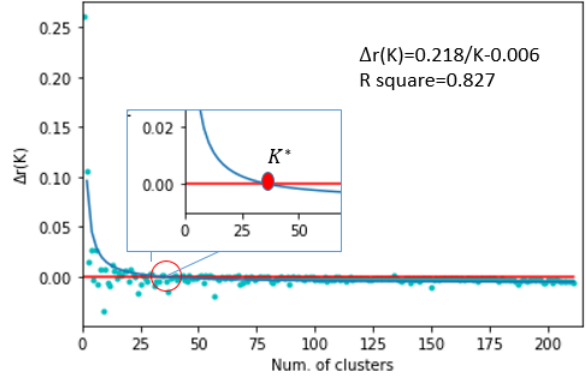

Fig. 4 The tendency of $\Delta r$ vs. K

### 4.5 Separable and inseparable data

According to the reference [4], there are two type of dataset based on the distribution of data point. They are separable data, which is distributed with certain bias, and inseparable data, which is evenly distributed. Accordingly, separable data rather than inseparable data are more detectable by clustering algorithms. Mathematically, separable data satisfy $\omega_{in}(K) \geq \omega_{out}(K)$ [4], where $\omega_{in}$ and $\omega_{out}$ are defined as follows.

$$\omega_{in}(K) = \frac{1}{K} \sum_{u \in g(k),\ v \in g(k)}^{G(K)} c_{uv} \tag{14}$$

$$\omega_{out}(K) = \frac{1}{K-1} \sum_{u \notin g(k),\ v \in g(k)}^{G(K)} c_{uv} \tag{15}$$

Accordingly, we generate these two types of data and test the performance of the ACCA. The experiment results in Fig. 5 indicate that the net benefit gap $r(K)$ resulting from a separable data presents a concave shape and the ACCA is able to stop at a local optimal number of clusters. But, the $r(K)$ resulting from the inseparable data is negative at most of the clustering solutions and does not present a concave shape. As the ACCA works on the inseparable data, it stops at the solution with a very small number of clusters such as one or two, which is just the preferable clustering solution for an inseparable data. These findings indicate that ACCA works efficiently for both separable and inseparable data. The experiments run on both Sioux Falls city network and Hardee Williston networks provide similar observations.

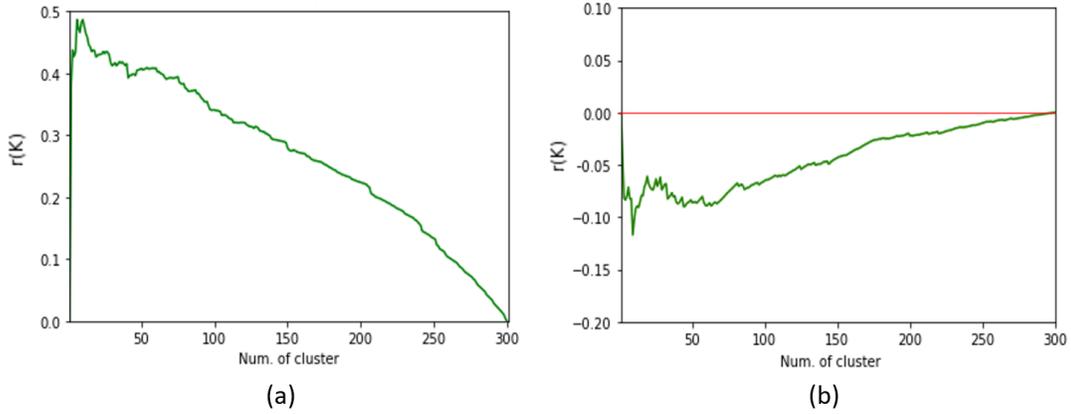

(a)                    (b)

Fig. 5 The r(K) wrt. the number of clusters for separable (a) and inseparable data (b) on Hardee network

### 4.6 Parallel Computation for the ACCA

It is noticed that the ACCA itself involves certain computation load, which will limit it applicability to online applications. This section analyzes its computation load in details and proposes parallel computation to address this difficulty. Specifically, after screening the computation loads of each step, this study recognizes that major computation load of the ACCA comes from the two procedures: building up the competition graph and searching for an optimal clustering solution. Below discuss their computation complexity and how to use parallel computation to improve its computing efficiency.

First of all, building up a competition graph needs to calculate the direct CP between every pair of travelers. Given CP is symmetric, it takes $0.5I^2$ iterations, even though Eq. (3) indicates that the calculation of each CP is a simple computation. Consequently, the complexity of this procedure is $O(I^2)$. On the other hand, we notice that the CP of each pair of travelers can be calculated independently. This feature provides the opportunity to further accelerate the computation process by implementing parallel computation.

We next discuss the computation load of calculating the indirect CP between every pair of travelers, which need to explore the clustering solution. According to Eq. (7), the indirect CP between any pair of travelers needs to solve an all-pairs shortest path problem. This study chooses a naïve scheme, which solves the shortest path for each pair of nodes by using Thorup's algorithm with computation complexity $O(I)$ [19] and then run it $I$ times for all pairs of nodes. This scheme will lead to a total complexity $O(I^2)$, which is better than some strategic algorithms such as Floyd-Warshall algorithm with computation load $O(I^3)$ [20]. Moreover, the indirect CP of each pair of travelers can be independently calculated. Using this naïve scheme will facilitate us to implement parallel computation to fast this computation more, thus it fits this study better. In addition, the Thorup's shortest path algorithm only works for undirected graph with positive integer weights, while the weights of the competition graph can be fractural numbers. To apply the Thorup's shortest algorithm, this study round all the weights to their closest integer. This transformation is viable since it does not violate the fundamental properties of indirect CP.

Last, we discuss the computing load of the procedure to explore a local optimal clustering solution. This process leads to a uncertain number of iteration. We, therefore, analyze the complexity of one iteration. The process to explore a clustering solution mainly includes two time-consuming operations. Referred to subsection 4.4.2, the first operation is the assignment of each node to its closest centroid with time complexity $O(KI)$. This operation can be implemented by parallel computation to speed up the process since the assignment for each node is independently conducted. The second main operation is to update the centroid for each cluster along with the new node assignment. To that, the algorithm can independently evaluate the concentration corresponding to each candidate centroid (i.e., each node) and then picks the one with minimum concentration as the new centroid for each cluster. As we discussed in subsection 4.4.2, its total time complexity is bounded by $O(I^2 + I)$. We can also use parallel computation to accelerate its computation process.

According to the above discussions, the flow chart of the ACCA implemented by the parallel computation is shown in Fig. 6. The computation performances with and without parallel computation of the ACCA is examined by numerical experiments in subsection 5.1.

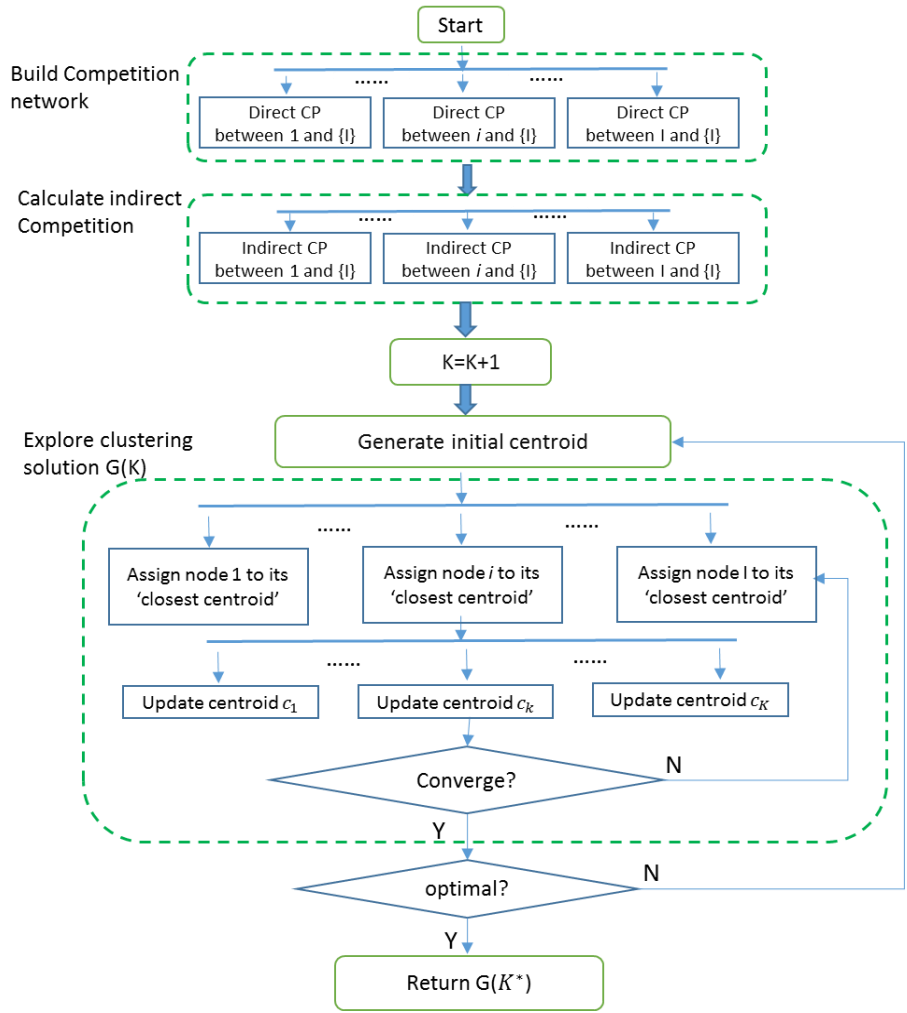

Fig. 6  Parallel computation flow chart of ACCA

## 4.7 Clustering based CRM

This section proposes a clustering based coordinated routing mechanism (CB-CRM) for coordinating the route choices of a large number of travelers en route, aiming to mitigate traffic congestion. Based on the understanding of the competition potential among travelers, the ACCA clusters these travelers into different coordination groups and then we independently implement the CRM [1] on each individual group and coordinate the route choices of travelers in each group. For the completeness, below provides the main procedure of CRM.

**Procedure of CRM**

**Initialization**: Predefined candidate routes and preference according to Eq. (16) for each traveler

**Iterations**:

If an equilibrium route choice among travelers are reached, exit;

Otherwise,

1) Each traveler updates its route choice preference according to real-time traffic condition factoring other vehicles' targeting route choice preferences

2) Each traveler proposes its route choice preference to real-time traffic information through connected vehicle environment

3) Information center synchronizes traffic information

**End**

A multinomial logit choice model is given in Eq. (16) below:

$$p_{t+1}^{\to i_h} = \frac{e^{-V_{i_h}}}{\sum_{i_h=1}^{i_H} e^{-V_{i_h}}} \tag{16}$$

where $V_{i_h} = \alpha^i + \beta^i C_t^{i_h}$ represents the utility of route $i_h$, and $C_t^{i_h}$ represents the travel time of route $i_h$.

It has been recognized that the inner-cluster CP in a coordinated group will potential promote the performance of the CRM in each group. However, the inter-cluster CP will likely interfere with the system performance. Our numerical experiments will exam how the inner-cluster and inter-cluster will affect the performance of the CB-CRM.

## 5. Numerical experiments

This study sets up numerical experiments to demonstrate the applicability of our approaches. Specifically, we would like to exam whether the ACCA can effectively cluster travelers according to the competition potential with acceptable computation efficiency. Moreover, the experiments will investigate the performance of the CB-CRM to demonstrate the value of integrating the ACCA into the real-time traffic management scheme. Accordingly, the experiments include three scenarios. They are (i) the IRM, by which each traveler independently choose the best route according to given real-time traffic information; (ii) the CRM, which considers all travelers en route as one coordination group and guide their route choices through a mixed strategy coordinated routing mechanism [1]; and (iii) the CB-CRM. The performance of the IRM, CRM, and CB-CRM is evaluated by the system travel time and they are denoted as $\sum \text{IRM}$, $\sum \text{CRM}$ and $\sum \text{CB-CRM}$ respectively.

The experiments are implemented on the Sioux Falls city network which includes 25 nodes and 76 links, and Hardee network which is consist of 44 nodes and 134 links. Link performance function is set as the BPR function. Travelers with different O-D pairs are randomly generated from multiple origin and destination zones to make it separable1. The candidate paths of each traveler are randomly assigned $i_H = 2$ or 3 and found by the k-shortest path algorithm [16] based on free flow traffic condition. The study runs multiple experiment cases by increasing the total travelers from 1,000 to 6,000 on Sioux Network (7000 on Hardee Network) with the step size 1,000, which makes the network traffic condition vary from LOS A to LOS F. Each case was run multiple times to avoid the effect of randomness. The experiments are implemented by MATLAB R2017a. on the HiperGator, the university supercomputer of UF. We run our experiments with requested computation resources 32 CPUs, and RAM: 1 GB per CPU. Below presents the main experimental results and observations.

### 5.1 Parallel computation for the ACCA

This study first exams how much parallel computation will improve the computation performance of the ACCA. The experiment results in Fig. 7 demonstrate that it can significantly improve the computation efficiency. As the overall number of travelers increases to 6000 on Sioux Network, the computation time of the ACCA is reduce from 2366 seconds to 101 seconds. This improvement makes the ACCA acceptable for online applications, such as the CB-CRM, when a large scale of travelers are involved.

---

1 The data, if a clustering algorithm is conducted, tends to have higher CP within the clusters than the CP between clusters[4]

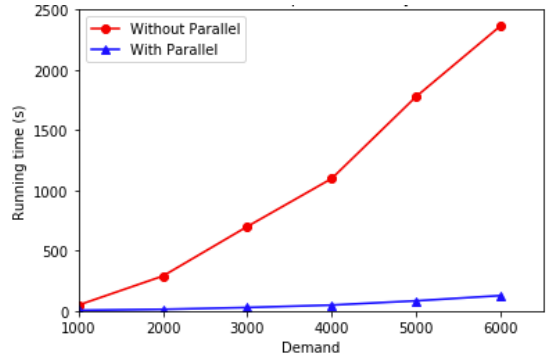

Fig. 7 Running time analysis for clustering procedure

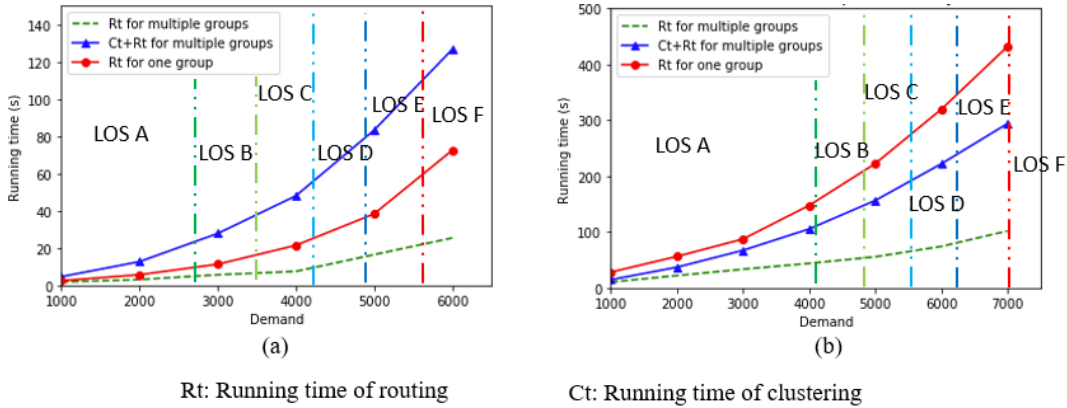

Rt: Running time of routing          Ct: Running time of clustering

Fig. 8 Running time analysis for Sioux Fall network (a) and Hardee network (b)

We next investigate the computation performance of the CB-CRM, which involves the computation time to complete the clustering by the ACCA (i.e. $C_t$, the running time of clustering), and to converge to an equilibrium in-vehicle routing decision among travelers in each cluster (i.e., $R_t$, the running time of routing) by the CRM. The experiment results for the Sioux Fall network and Hardee network are shown in the Fig. 8 (a) and (b) respectively.

First of all, both Fig. 8 (a) and (b) indicate that the running time of $R_t$ (the green dash line) in the scenario of the CB-CRM with multiple small condition groups is smaller than $R_t$ (the red solid line with dot markers) in the scenario of the CRM implemented on the entire large traveler group. Thus, the benefit of using the ACCA to splits the entire traveler group into small clusters is first demonstrated by improving the computation efficiency of routing. Next, we exam the total computation load of the CB-CRM and CRM (i.e., ($C_t$+$R_t$) for the CB-CRM vs. $R_t$ for the CRM). Fig. 8 (a). demonstrates that the total computing time of the CB-CRM (the blue line with triangle markers) is higher than that of the CRM on Sioux Fall network, even though they are both applicable for the online application (the maximum computation time less than 127 seconds and 73 seconds respectively under the LOS F case). However, Fig. 8 (b) demonstrates reverse results on Hardee network. It shows that the computation time of the CB-CRM (the blue line with triangle markers) is less than that of the CRM under different traffic condition cases. Moreover, the merit becomes more apparent as the traffic becomes more congested. Fig. 8 (a) and Fig. 8 (b) together demonstrate that the ACCA is only worthwhile to be integrated with the CRM for coordinating a large scale of travelers in a big network since the reduction of the running time for routing in an individual coordinated group can compensate the clustering computation time. Furthermore, the CRM requires more communication between travelers and information center since a large size of coordination group tends to require more iterations for the routing mechanism to converge (shown in the following experiments). Thus, the CRM may perform even worse than the CB-CRM in practice since it has multiple smaller groups and requires much less communication time. Besides the running time of the CB-CRM can be further reduced by using more CPUs. All these merits of the CB-CRM indicate that it is more applicable than the CRM in practice.

Last, except the total computation performance, the CB-CRM perform similarly on the Sioux Fall network and Hardee network. Thus, we present the results on the Hardee network in the following discussions.

### 5.2 Validating the direct CP formulations

Recall that the formulations in Eq. (1)-(5) are developed to evaluate the direct CP between two travelers on the usage of route resources in a traffic network. Considering the direct CP will affect the measurement of the indirect CP between two travelers and among multiple travelers, our experiments first test the performance of the CB-CRM as different direct CP formulations are implemented.

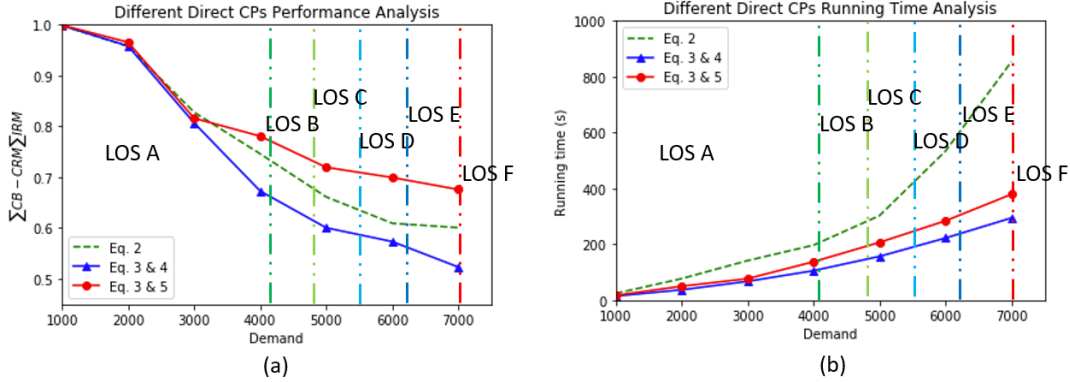

Fig. 9 System performance (a) and computation (b) analysis wrt. various CP measures

The results in Fig. 9 (a) indicate that the system travel time of the CB-CRM's integrating different formulations performs similarly under free-flow traffic condition (LOS A). But, the CB-CRM using temporal-spatial formulation (Eq. (3) together with Eq.(4)) results in better system performance than using Eq. (2), which only counts spatial overlap, and Eq. (3) together with (5), which counts the uncertainty of travel time, as traffic gets congested. More exactly, using Eq. (3) together with Eq. (5) leads to a worst system performance. These results indicate that the uncertainty of travel time is not significant to the CB-CRM. This study senses the reason is that the CRM doesn't count the uncertainty of the arrival time to individual links, thus the formulation of Eq. (3) together with Eq. (5) don't fit the CRM very much. On the other hand, the CB-CRM using the formulation of Eq. (3) together with Eq.(4) presents the best computation performance too (see the blue solid line with triangle markers in Fig. 9 (b)). The reason is that the constructed CP network is much denser than using Eq. (2) since just considering spatial overlap causes more unnecessary CP between travelers, which bring up the computation load of calculation of indirect CP between any two travelers; in addition, Eq. (3) together with (5) causes more computation load than Eq. (3) together with Eq.(4) since it considers the uncertainty of travel time as well. Overall, the CB-CRM integrating Eq. (3) together with Eq. (4) outperforms other formulations in both the system performance and computation efficiency. It is thus applied by the study in the rest of the experiments.

### 5.3 Validating the optimality of the ACCA

This study further validates whether the optimal clustering solution of the ACCA will lead to local optimal system performance of the CB-CRM scenario, considering the clustering solution as the variable. To do that, the experiments run the ACCA without stopping at the optimal solution so that we can obtain multiple clustering solutions. Next, we check the system travel time ($\Sigma$) as the CB-CRM is implemented by using different clustering solutions. The results in Fig. 10 demonstrate that the CB-CRM will lead to better system performance than the IRM (i.e., $\sum$ CB-CRM / $\sum$ IRM < 1) under most of the clustering solutions. However, the vantage gets weaker as the number of the clusters increases. More importantly, there are some points, where the system performance of the CB-CRM scenario drops dramatically and the benefit-cost gap $r(K)$ reaches to a local optimal such as K=7 and K=14. Note that the ACCA prefers stopping at the first local optimal $r(K)$. According to this criterion, the ACCA will stop at $K = 7$. Therefore, we can claim that the optimal clustering solution of the ACCA will ensure a local optimal system performance too.

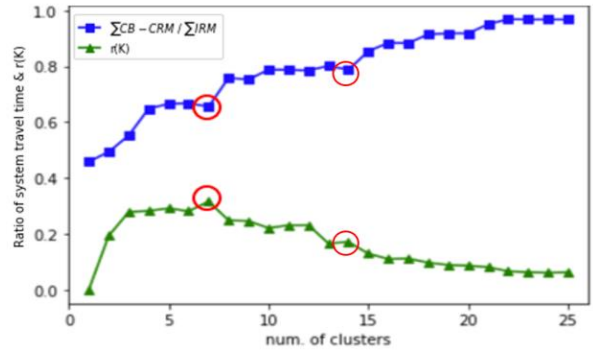

Fig. 10 System performance and r(K) wrt. the number of clusters

### 5.4 The ACCA vs random clustering approach

This section further justifies the effectiveness of the ACCA by comparing the system travel time ($\Sigma$) under the scenarios of the CB-CRM and the RCA-CRM, which respectively implement independent CRM on the clustering solutions obtained from the ACCA and from a random clustering approach (RCA). The RCA will evenly and randomly assign travelers to different clusters. Fig. 11 illustrates the results regarding the system travel time in the case of 7,000 travelers. Note that previous study [1] already showed that the CRM leads to less network level traffic congestion than the IRM. It is also confirmed by this study in Fig. 12 (a), in which both $\Sigma CRM/\Sigma IRM$ and $\sum CB\text{-}CRM / \sum IRM$ are less than one. The experiment results in Fig. 11 show that the ACCA is a more effective clustering approach than the RCA to support the merit of the CRM to IRM since the CB-CRM leads to a better system performance than the RCA-CRM does (i.e., $\sum CB\text{-}CRM / \sum IRM < \sum RCA\text{-}CRM / \sum IRM$ in Fig. 11). The study runs multiple experiments including the number of the travelers increasing from 1,000 to 7,000 with the step size 1000. We observed the similar results shown in Fig. 11.

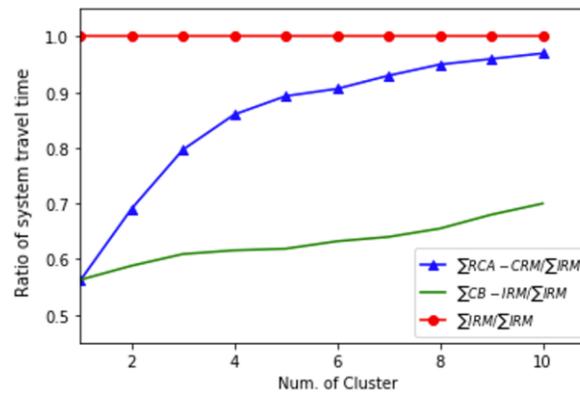

Fig. 11 System performance of the CB-CRM and RCA-CRM wrt. the number of clusters

*5.5 The CB-CRM for separable data*

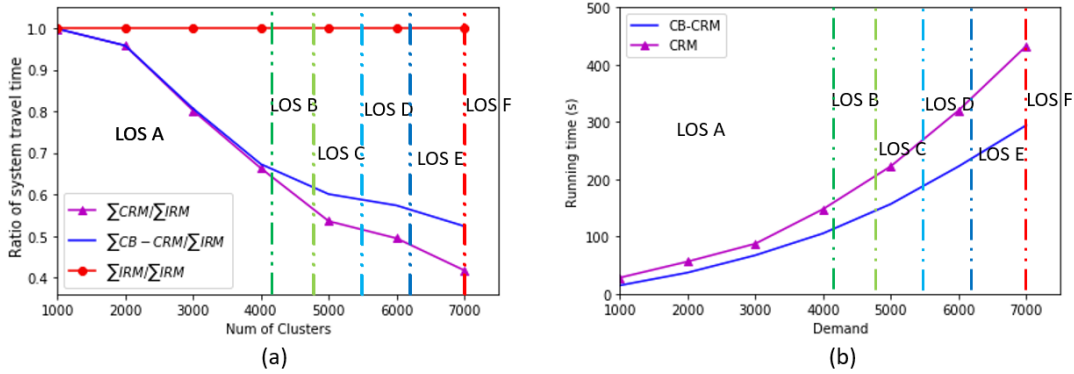

Fig. 12 System performance (a) and computation analysis (b) of the CB-CRM for separable travelers

This study first exams the performance of implementing the CB-CRM under separable data. Namely, the travelers are distributed in the network with a bias. The results in Fig. 12 demonstrate that the CB-CRM results in almost the same system performance as the CRM under LOS A. However, as traffic gets more congested such as LOS B to LOS F, the CB-CRM performs slightly worse than the CRM but still much better than the IRM. On the other hand, the CB-CRM needs much less computation resource than the CRM. For example, under the case of 7,000 travelers, the system travel time under the scenario of the CB-CRM is 17% more than that under the CRM scenario. However, it is still about 48% less than that under the scenario of the IRM. On the other hand, we can see that the running time of the CB-CRM scenario is 31% less than that of the CRM scenario. Therefore, we conclude that the CB-CRM helps reduce the computation load with a minor loss in the system performance as compared to the CRM under the separable data.

*5.6 The CRM for inseparable data*

This study next wants to investigate the performance of the CB-CRM for inseparable data. Namely, travelers are evenly distributed over a network. Fig. 13 shows the experiment results for the case with 7,000 travelers. It is illustrated that the CB-CRM is still beneficial to reduce system travel time from the IRM when a large scale of travelers involved in the routing coordination, even though it performs slightly worse than the CRM due to the interference of the inter-cluster competition. In addition, the CB-CRM presents the vantage in computation efficiency as compared to the CRM. More exactly, we can see when K = 4, the CB-CRM loses 5% of the merit in system travel time but gain around 44% of computation time as compared to CRM. Therefore, we conclude that the CB-CRM is still applicable and beneficial for inseparable data. Note that this study also run the experiments with the number of travelers change from 1000 to 7000 and we get similar observations.

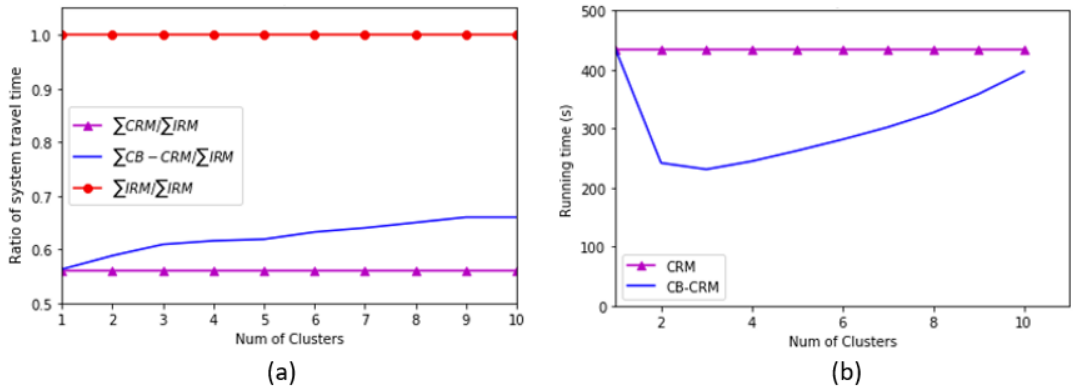

Fig. 13 System performance (a) and computation analysis (b) for inseparable data

### 5.7 The CB-CRM under different penetration

This study further considers that not all travelers in the network will opt in the coordinated routing mechanism, and then investigates the merits of the CB-CRM under different percentages (i.e., the percentage of the travelers opts in the coordinated routing mechanism). To do that, this study run experiments in which the number of total travelers varying from 1,000 to 7,000 to cover LOS from A to F. Under each case, we vary the percentage of the travelers following the CB-CRM (or the CRM) from 10% to 100% with step size 10%. The rest of the travelers follows the IRM. The vantage of the CB-CRM to the CRM is evaluated by comparing its system travel time (ΣCB-CRM) to the corresponding values of the scenarios of the CRM (ΣCRM) and the IRM (ΣIRM). The results in Fig. 14 (a) demonstrate that the advantage of the CB-CRM to the CRM lose more and more as the penetration and congestion level increase. This is because more inter-cluster competition presents and degrades the function of the CB-CRM as more travelers opt in the coordinated routing mechanism. However, Fig. 14 (b) shows that the CB-CRM presents more advantage to the IRM as the penetration and congestion level increase. Especially, when penetration is greater than 70% and LOS is worse than C, the CB-CRM improves the system performance by more than 30%. Moreover, as we already knew that the CB-CRM saves the computation load from the CRM. These good qualities ensure the advantages of implementing the CB-CRM on a large network.

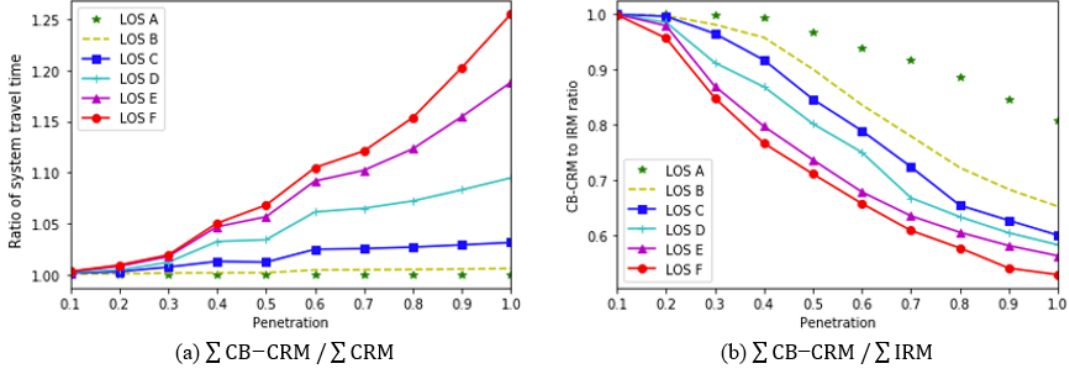

(a) $\sum$ CB-CRM / $\sum$ CRM    (b) $\sum$ CB-CRM / $\sum$ IRM

Fig. 14 System performance analysis wrt. penetration

### 5.8 Discrete choice model Experiments

Last, this study noticed that the choice model in Eq. (16) does not count the effect of the overlapping among the candidate routes of a traveler. We thus conduct experiments to test the performance of the CB-CRM but using the C-Logit route choice model in Eq. (17), where $\hat{\beta} = 1, \gamma = 2$. The results in Fig. 15 indicate a similar tendency to the results in Fig. 14, in which Eq. (16) is used as the choice model, but presents better system performance. The observation further reinforces the value to consider the competition potential among travelers for mitigating network level traffic congestions.

$$p_{t+1}^{\rightarrow i_h} = \frac{e^{-\hat{V}_{i_h}}}{\sum_{i_h=1}^{i_H} e^{-\hat{V}_{i_h}}}$$

$$\hat{V}_{i_h} = V_{i_h} - CF_{i_h}$$

$$(17)$$

$$CF_{i_h} = \hat{\beta}\ln\sum_{j_h \in i_H}\left[\frac{\sum_{a \in a_{i_h} \cap a_{j_h}} \bar{t}_a o_a}{\left(\sum_{a \in a_{i_h}} \bar{t}_a\right)^{\frac{1}{2}}\left(\sum_{a \in a_{j_h}} \bar{t}_a\right)^{\frac{1}{2}}}\right]^{\gamma}$$

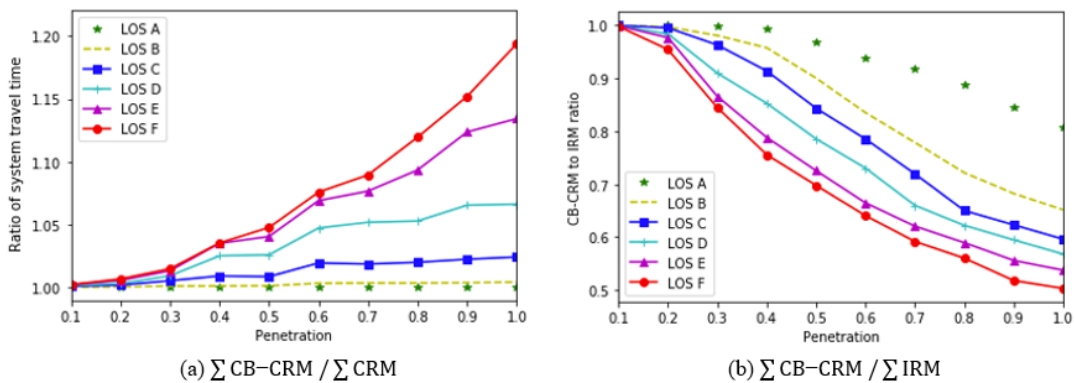

(a) $\sum$ CB−CRM / $\sum$ CRM　　　(b) $\sum$ CB−CRM / $\sum$ IRM

Fig. 15 System performance wrt. the penetration (C-Logit model)

## 6. Conclusion and future work

It is noticed that the competition potential among travelers driving en route on the usage of common traffic network resources such as routes, road capacity, parking lots, etc. provides a great potential for traffic operators to implement online coordinated traffic management and control schemes so that we can mitigate network-level traffic congestion. However, very few studies work on this research issue and we're still lack of quantitative approaches to better understand and further take advantage of these phenomena. Motivated by this view, this research seeks to partially make up this technical gap. Specifically, this study focuses on the competition potential among the travelers on the usage of route resources on the network and seek to well coordinate their online route choices to mitigate traffic congestion. Correspondingly, the research efforts in this paper contribute to the following methodologies.

First of all, this study thinks that the temporal-spatial overlap among the candidate routes of two travelers represents the potential that these two travelers will compete for the usage of these routes en route. Accordingly, we develop several formulations to quantify this direct competition potential between two travelers, respectively taking account of spatial, temporal-spatial route overlap, the uncertainty of link travel time or their combined effect. Built upon that, this study establishes the competition network in Fig. 2, from which this study quantifies the indirect competition potential between any two travelers by the shortest path approach; and further measures the competition potential among a group of travelers by the concentration. Moreover, we develop the ACCA to cluster a large scale of travelers into multiple coordination groups. The algorithm design creates a marginal net benefit criterion to ensure a local optimal clustering solution, which well balances the tradeoff between the size of each cluster and the number of clusters (i.e., inner-cluster vs. inter-cluster competition potential) so that coordinated routing mechanisms (such as the CB-CRM) can be well implemented, pursuing both system performance improvement and reasonable computation efficiency. The optimality of the local optimal clustering solution is ensured by a statistical discussion.

The numerical experiments conducted on Sioux Falls and Hardee city networks indicate several good features of the ACCA. First of all, the computation load of ACCA can be significantly reduced through the implementation of parallel computation. The ACCA works more effectively than the RCA to group travelers into a cluster with strong inner-cluster competition potential. The stopping criterion of the ACCA ensures local optimal clustering solution which corresponds to a local optimal system performance of the CB-CRM. Moreover, the CB-CRM is able to greatly improve the computation efficiency of the routing mechanism, while keeps the satisfying system performance as compared to implementing the CRM considering all travelers as one coordination group. However, due to the computation load of the ACCA, the CB-CRM is more applicable for a large network like Hardee rather than a small network like Sioux Fall. The CB-CRM leads to more system travel time reduction from the IRM as the penetration and congestion level increase. The ACCA and the CB-CRM work efficiently for both separable and inseparable data. The CB-CRM employing the C-Logit choice model will lead to a better system performance than using Multimode Logit model. The research conducted in this paper can be further extended to other applications such as parking and riding sharing, in which users also present the potential competition or cooperation. Coordinated management is applicable for a better system performance. We will explore these potential studies in our future work.

**Acknowledgements**

This research is partially supported by the National Science Foundation awards CMMI-1436786 and CMMI-1554559.

**Reference**

[1] Du, Lili, Lanshan Han, and Xiang-Yang Li. "Distributed coordinated in-vehicle online routing using mixed-strategy congestion game." *Transportation Research Part B: Methodological* 67 (2014): 1-17

[2] Noland, Robert B., and John W. Polak. "Travel time variability: a review of theoretical and empirical issues." *Transport reviews* 22.1 (2002): 39-54.

[3] Hartigan, John A., and Manchek A. Wong. "Algorithm AS 136: A k-means clustering algorithm." *Journal of the Royal Statistical Society. Series C (Applied Statistics)* 28.1 (1979): 100-108.

[4] Von Luxburg, Ulrike. "A tutorial on spectral clustering." *Statistics and computing* 17.4 (2007): 395-416.

[5] Grygorash, Oleksandr, Yan Zhou, and Zach Jorgensen. "Minimum spanning tree based clustering algorithms." *Tools with Artificial Intelligence, 2006. ICTAI'06. 18th IEEE International Conference on*. IEEE, 2006.

[6] Bouguettaya, Athman, et al. "Efficient agglomerative hierarchical clustering." *Expert Systems with Applications* 42.5 (2015): 2785-2797.

[7] Ester, Martin, et al. "A density-based algorithm for discovering clusters in large spatial databases with noise." *Kdd*. Vol. 96. No. 34. 1996.

[8] Wang, Kaijun, et al. "Adaptive affinity propagation clustering." *arXiv preprint arXiv*: 0805.1096 (2008).

[9] Milligan, Glenn W., and Martha C. Cooper. "An examination of procedures for determining the number of clusters in a data set." *Psychometrika* 50.2 (1985): 159-179.

[10] Kaufman, Leonard, and Peter J. Rousseeuw. "Partitioning around medoids (program pam)." *Finding groups in data: an introduction to cluster analysis* (1990): 68-125.

[11] Fortunato, Santo, and Darko Hric. "Community detection in networks: A user guide." *Physics Reports* 659 (2016): 1-44.

[12] Park, Hae-Sang, and Chi-Hyuck Jun. "A simple and fast algorithm for K-medoids clustering." *Expert systems with applications* 36.2 (2009): 3336-3341.

[13] Miller, Harvey J., and Shih-Lung Shaw. *Geographic information systems for transportation: principles and applications*. Oxford University Press on Demand, 2001.

[14] Ben-Akiva, Moshe, Andre de Palma, and Isam Kaysi. "The impact of predictive information on guidance efficiency: An analytical approach." *Advanced methods in transportation analysis*. Springer, Berlin, Heidelberg, 1996. 413-432.

[15] Dijkstra, Edsger W. "A note on two problems in connexion with graphs." *Numerische mathematik* 1.1 (1959): 269-271.

[16] Yen, Jin Y. "Finding the k shortest loopless paths in a network." *management Science* 17.11 (1971): 712-716.

[17] Peizer, David B., and John W. Pratt. "A normal approximation for binomial, F, beta, and other common, related tail probabilities, I." *Journal of the American Statistical Association* 63.324 (1968): 1416-1456.

[18] Peters, Tim. "Timsort." (2002). http://svn.python.org/projects/python/trunk/Objects/listsort.txt

[19] Thorup, Mikkel. "Undirected single-source shortest paths with positive integer weights in linear time." *Journal of the ACM (JACM)* 46.3 (1999): 362-394.

[20] Cormen, Thomas H., et al. *Introduction to algorithms*. MIT press, 2009.